\documentstyle [12pt,epsf]{article}

\input{epsf}
\begin{document}
\begin{titlepage}
\begin{center}

 \vspace{-0.7in}

{\large \bf A Continuously Observed Two-level System \\ Interacting with a Vacuum Field}\\
 \vspace{.3in}{\large\em
R. Kullock\,\footnotemark[1]and N. F. Svaiter
\footnotemark[2]}\\
\vspace{.3in}
 Centro Brasileiro de Pesquisas F\'{\i}sicas,\\
 Rua Dr. Xavier Sigaud 150,\\
 22290-180, Rio de Janeiro, RJ Brazil \\

\subsection*{\\Abstract}
\end{center}

\baselineskip .18in

A discussion of the quantum Zeno effect and paradox is given. The
quantum Zeno paradox claims that a continuously observed system,
prepared in a state which is not an eigenstate of the Hamiltonian
operator, never decays. To recover the classical behavior of
unstable systems we consider a two-level system interacting with a
Bose field, respectively prepared in the excited state and in the
Poincar\'e invariant vacuum state. Using time-dependent
perturbation theory, we evaluate for a finite time interval the
probability of spontaneous decay of the two-level system. Using
the standard argument to obtain the quantum Zeno paradox, we
consider $N$ measurements where $N \rightarrow \infty$ and we
obtain that the non-decay probability law is a pure exponential,
therefore recovering the classical behavior.

\footnotetext[1]{e-mail:\,\,ricardokl@cbpf.br}
\footnotetext[2]{e-mail:\,\,nfuxsvai@cbpf.br}

PACS numbers:03.70+k,04.62.+v

\end{titlepage}
\newpage\baselineskip .18in
\section{Introduction}

In the present paper we are interested to show that a continuously
observed quantum system has a classical time evolution behavior if
it interacts with a vacuum field. Being more specific, we study
the time evolution of a two-level system, i.e., a qubit,
interacting with a Bose field prepared in the Poincar\'e invariant
vacuum state, showing that the non-decay probability has a
exponential behavior.

Unstable systems in quantum mechanics have been the subject of
many investigations since the origin of this formalism, and have a
long story starting with the seminal papers of Gamow \cite{gamow}
and Wigner \cite{wigner}. The temporal evolution of these quantum
systems can be roughly divided in three different behaviors. A
gaussian-like behavior at short times, an exponential decay at
intermediate times and finally a power-law decay at long times
\cite{fonda} \cite{peres} \cite{namiki}. The gaussian behavior for
short times is the key point of our discussions bellow, since it
leads under general physical conditions to the inhibition of the
decay of unstable quantum mechanical systems.

This result was obtained by Misra and Sudarshan \cite{misra}
\cite{misra2} using the von Neumann description of measurement
processes \cite{von}. These authors proved that the realization of
many successive measurements dramatically slow down the evolution
of an unstable system. In the limit of continuous observations,
the temporal evolution of a quantum system can be frozen, defining
respectively the quantum Zeno effect and paradox, after the
paradox given by the Greek philosopher Zeno \cite{plato}.
Actually, many years before, a close related result to the one
obtained by Misra and Sudarshan was achieved by Khalfin
\cite{khalfin}, where a proof of the deviation from the
exponential decay-law for large times was given.

The first conceptual question in the Misra and Sudarshan
construction is the following: we may ask whether is possible to
realize this limit of continuous observation. Although some
authors argue that this limit of continuous observation is not
physical and must be regarded as a mathematical idealization
\cite{home2} \cite{home1} \cite{nakazato}, we prefer to discuss
this delicate issue later.

As has been stressed by many authors \cite{caldeira} \cite{joos}
\cite{li} \cite{grabert} \cite{merchede} \cite{puri}
\cite{breuer}, any quantum mechanical system which we are
interested to investigate its behavior on time interacts with the
surroundings. In other words, in Nature we deal actually with open
systems which are influenced by the surrounding world through
exchange of energy, or in a more abstract way, information. These
considerations motivate us to investigate unstable systems which
are continuously observed in a finite time interval, including a
vacuum field in the problem. The two-level system, prepared in a
eigenstate of its free Hamiltonian, can make a transition to the
lower eigenstate driven by the vacuum field, which acts as a
reservoir. This is a development of an old idea. In the derivation
of Planck's radiation law, Einstein introduced the idea of
spontaneous emission, where a system makes a transition to a lower
eigenstate without external stimulation \cite{einstein}.

In quantum mechanics, in treating arbitrary systems the time
evolution of closed systems is described by one-parameter group of
unitary operators, and the equation of motion of such systems are
symmetric to time reversal. When a small system interacts with an
reservoir, which is characterized by an infinitely large number of
degrees of freedom, the time evolution of the small system can not
be represented in terms of unitary Hamiltonian dynamics and we
call it an open quantum system. There are a variety of theoretical
models of reservoirs. One situation is when the system $S$ is
coupled to an infinite number of harmonic oscillators. In this
situation there are two kinds of reservoir of common interest. The
first one is a thermal reservoir, where we assume that the
harmonic oscillators are in thermal equilibrium at temperature
$\beta^{-1}$. The second one is a squeezed reservoir. The specific
system-reservoir model which is appropriate for the study of
several interesting situations is when the harmonic oscillator
bath is constituted by a Bose field in free space.

The aim of the paper is to recover the classical time evolution
behavior for continuously observed systems. The situation that we
are interested to study is when the two-level system is prepared
in the excited state and it is interacting with a Bose field
prepared in the Poincar\'e invariant vacuum state. Using
perturbation theory we compute the probability of decay evaluated
in a finite time interval.

These calculations are not new in the literature. Discussing model
detectors, Svaiter and Svaiter \cite{ss} \cite{errata} assumed a
weak-coupling between a two-level system and a Bose massless
field. These authors evaluate the transition rates of the
two-level system in different kinematic situations without use of
the rotating-wave approximation \cite{ja2} \cite{dekker}. Further,
Ford and collaborators \cite{fs}, using the same model, assumed
the presence of one or two infinite perfectly reflecting plates
(mirrors). They show how these mirrors, which change the vacuum
fluctuations associated to the Bose field, influence radiative
processes at zero temperature. They also evaluate the probability
per unit time of spontaneous emission at finite temperature.
Radiative processes of atoms in waveguides and cavities also have
been investigated by many authors. See for example the Ref.
\cite{meschede}. In the Refs. \cite{vitorio} \cite{paola} one of
the authors continue to investigate radiative processes associated
to the Unruh-DeWitt detector \cite{unruh} \cite{Dewitt}
\cite{davies}, in interaction with a massless scalar field. Being
more precise, in Ref. \cite{paola}, it was calculated the
detector's excitation rate when it is uniformly rotating around
some fixed point, when the scalar field is prepared in the
Poicar\'e invariant vacuum state, and also when the detector is
inertial and the field is prepared in the Trocheries-Takeno vacuum
state \cite{vi1} \cite{vi2}. These two response functions allow to
the authors to present questions analogous to those discussed by
Mach in the Newton's bucket experiment in a quantum mechanical
level.

This paper is organized as follows. In Section II we briefly
discuss the theory of the classical and quantum mechanical decays
and the quantum Zeno paradox. In Section III, assuming that the
two-level system interact with a Bose field prepared in the
Poincar\'e invariant vacuum state, the probability of decay of the
two-level system is evaluated in a finite time interval. In
Section IV, using the same arguments as in the quantum Zeno
paradox, we obtain that in the case of continuous observations the
non-decay probability law is exponential for all times.
Conclusions are given in Section V. In the Appendix we discuss
qubit-boson field interaction Hamiltonians. In the paper we use
$k_{B}=c=\hbar=1$

\section{The classical and quantum mechanical decays
and the quantum Zeno paradox}\

To introduce probability in classical physics we have to make use
of a huge number of identical prepared systems. The classical
theory of decay is quite simple and is based on the assumption
that unstable systems have a certain probability of decay. The
basic features of this simple model is that we assume a Markovian
approximation. Therefore this probability does not depend on the
past history of the unstable system. Let us assume $N$ unstable
systems, and that the decay probability per unit time be a
constant that we call $\Gamma$. For simplicity $\Gamma$ is
characteristic of the system and also does not depend on the total
number of unstable systems nor on the environment surrounding
them. Let us define the number of unstable systems at time $t$ by
$N(t)$. Therefore the number of systems that will decay in the
infinitesimal interval of time $dt$ in $dN(t)$. Consequently we
have
\begin{equation}
-dN(t) = N \, \Gamma \, dt .
\label{cdecay1}
\end{equation}
Defining the inverse of $\Gamma$, i.e., the lifetime of the
unstable system by  $\tau_{E} \, \, (\Gamma =
\frac{1}{\tau_{E}})$, the number of unstable systems at a generic
time $t$ is
\begin{equation}
N(t) = N(0) \,\exp{\left(-\frac{t}{\tau_{E}}\right)}  ,
\label{cdecay2}
\end{equation}
where $N(0)$ is the number of unstable systems at the beginning of
our observation, i.e., $t=0$. One define the non-decay classical
probability $P_{class}(t)$ as
\begin{equation}
P_{class}(t) = \frac{N(t)}{N(0)} =
\exp{\left(-\frac{t}{\tau_{E}}\right)}. \label{cdecay3}
\end{equation}
For short times $(t<<\tau_{E})$ we can write
\begin{equation}
P_{class}(t) = 1 - \frac{t}{\tau_{E}} + ...
\label{cdecay4}
\end{equation}
Note that we are excluding cooperative effects, therefore $\Gamma$
and also $P(t)$ are environment-independent. The solution given by
the Eq.(\ref{cdecay3}) has a dissipative behavior and is a
fundamental law that gives the classical behavior of unstable
systems, as in experimental nuclear physics, for instance.

In quantum mechanics we introduce probability even working with a
single system. A quantum mechanical treatment of the same problem
give us a short and large time behaviors which are in disagreement
with the exponential law obtained in Eq.(\ref{cdecay3}). Let us
first discuss the deviation from the exponential decay law for
large times. We are following the arguments presented in the Ref.
\cite{misra2}. Let us assume a quantum system with a set of
observables, i.e., operators which commute with the Hamiltonian of
the system. The Hamiltonian has a complete set of eigenstates,
which is a basis of the Hilbert space, therefore every state
vector of the system can be expressed in terms of it. For
simplicity we assume that there is only one unstable state that we
represent by $| \, a \, \rangle$, which is orthogonal to the bound
states of the Hamiltonian operator $H$. Since the dynamic is
time-translational invariant, the unitary operator
$U(t_{2}-t_{1})$ propagates the system from $t_{1}$ to $t_{2}$.
Using the self-adjoint operator $H$ of the system, the dynamics is
defined by the unitary operator $U(t)=e^{-iHt}$. Suppose that we
are studying the temporal evolution of the system after $t=0$.
Let us define the spectral projection of the Hamiltonian operator
by
\begin{equation}
H=\int\,d\lambda\,\lambda | \,\lambda \, \rangle\,\langle
\,\lambda \, |.
\label{n1}
\end{equation}
The energy distribution function of the state $| \, a \, \rangle$
or the probability that the energy of the unstable state $| \, a
\, \rangle$ lies in the interval $[E,E+dE]$ is given by
\begin{equation}
\int_{E}^{E+dE}\, \langle\,a|\,\lambda
\rangle\,\langle\,\lambda|\,a \rangle\,d\lambda. \label{n2}
\end{equation}
The non-decay probability at the time $t$ is defined by
$P_{quant}(t)$. Therefore, using the standard interpretation of
quantum mechanics, we have that
\begin{equation}
P_{quant}(t) = |\langle \, a \, | \, e^{-iHt} \, | \, a \,
\rangle|^2,
 \label{qdecay1}
\end{equation}
where the decay probability is given $[1-P_{quant}(t)]$. Let us
study the non-decay amplitude. It is given by
\begin{equation}
\langle \, a \, | \, e^{-iHt} \, | \, a \, \rangle=\int
d\lambda\,e^{-i\lambda t} \, \langle\,a|\,\lambda
\rangle\,\langle\,\lambda|\,a \rangle\,d\lambda.
\label{n2}
\end{equation}
If $\langle\,\lambda|\,a \rangle=0$, for $\lambda < 0$, i.e., the
spectrum of the Hamiltonian operator $H$ is bounded from bellow,
then when $t\rightarrow \infty$ the quantity $P_{quant}(t)$
decreases to zero less rapidly than any exponential of the form
$e^{-\sigma t}$. Therefore we have a deviation from the
exponential decay law at large times. Also, the treatment for the
same problem for short time give us a short time behavior which is
quadratic and therefore in disagreement with the exponential law
obtained in Eq.(\ref{cdecay3}).

Let us assume again a quantum system with an Hamiltonian operator
$H$ with a complete set of eigenstates denoted by $|\,i\,\rangle$
$(i=1,2,3,...)$. If we prepare the system in a normalized state
$|\,a\,\rangle$ which is not an eigenstate of $H$, it is possible
to show that the non-decay probability at short times is of the
gaussian type.
A short time expansion using the Eq.(\ref{qdecay1}) yields
\begin{equation}
P_{quant}(t) = 1- \frac{t^2}{\tau_{z}^2} + ... ,
\label{qdecay2}
\end{equation}
where the quantity $\tau_{z}^{-1} = [ \langle \, a \, | \, H^2 \,
| \, a \, \rangle - \langle \, a \, | \, H \, | \, a \, \rangle^2
]^{\frac{1}{2}}$ is the inverse of the characteristic time of the
gaussian evolution. The crucial feature of this approximation is
that the non-decay probability after a short observation time $t$
is quadratic. The quantity $\tau_{z}$ is also called the Zeno
time.

Let us assume that the quantum measurement occurs instantaneously.
We also assume that is possible to perform infinitely many
measurements in a given finite interval. Suppose that we perform
$N$ measurements at equal time interval which satisfies $T = N
\Delta\tau$. In each measurement we observe that the system stays
in the initial state $|\, a \, \rangle$ which was defined before.
The probability of observing the initial state at the final time
$T$ after $N$ measurements reads
\begin{equation}
P^{(N)}_{quant}(T) = \left[ P_{quant} \left( \frac{T}{N} \right)
\right]^N. \label{qdecay3}
\end{equation}
Substituting Eq.(\ref{qdecay2}) in Eq.(\ref{qdecay3}) we have
\begin{equation}
P^{(N)}_{quant}(T) \approx \left[ 1- \frac{1}{\tau_{z}^2} \left(
\frac{T}{N} \right)^2 \right]^N . \label{qdecay4}
\end{equation}
For very large $N$ we get
\begin{equation}
P^{(N)}_{quant}(T) \approx \, \exp{\left( -\frac{T^2}{\tau_{z}^2
N} \right)}, \label{qdecay5}
\end{equation}
and repeated observations slow-down the evolution of the unstable
system and increase the probability that the system remains in the
initial state at $T$. If we are able to set $N \rightarrow \infty$
one obtains
\begin{equation}
\lim_{N \rightarrow \infty} P_{quant}^{(N)}(T) \approx \, \lim_{N
\rightarrow \infty} \exp{\left(-\frac{T^2}{\tau_{z}^2 N}\right)} =
1. \label{qdecay6}
\end{equation}
This is a very simple derivation of the quantum Zeno paradox. The
unstable quantum system becomes stable if we perform infinitely
continuous measurements.

There are many physical assumptions that we have to make to obtain
this effect. Many authors claim that the limit of infinite
measurements is non-physical, and it is in contradiction with the
Heiseberg uncertainty principle \cite{ghosh} \cite{peti}. We leave
open these questions right now and we shall come back to this
important issue when we discuss the interpretation of time-energy
uncertainty relations \cite{landau1} \cite{landau2} \cite{fock}
\cite{aharonov} \cite{bohm}.

In the next Section we are interested in the quantum measurement
of a single object interactng with the vacuum modes and how to
evaluate the probability of transition in a finite observation
time. Since we are assuming a weak-coupling between the two-level
system and the environment, the probability of decay of the
two-level system is computed using first-order approximation in
perturbation theory.

\section{The probability of decay evaluated for a finite time interval }\

In this section we are interested in computing radiative processes
of a quantum two-level system, interacting with a vacuum field.
For simplicity we will use the following notation. The two energy
levels of the two-level system, i.e., the ground and excited
energy levels, are given by $\omega_{g}$ and $\omega_{e}$
$(\omega_{e}-\omega_{g}=\omega\,>\,0)$, with eigenstates of the
free two-level system Hamiltonian $|\,g\,\rangle$ and
$|\,e\,\rangle$, respectively. We are assuming a non-zero monopole
matrix element between these two states and we can assume that the
diagonal elements of the monopole operator vanish.

As we discuss in the appendix, the coupling between the massless
scalar field and the two-level system is given by a monopole
interaction Hamiltonian, i.e.,
\begin{equation}
H_{I}=\lambda\,m(\tau)\,\varphi(x(\tau)), \label{34}
\end{equation}
where $m(\tau)$ is the monopole operator of the two-level system,
$\varphi(x(\tau))$ is the scalar field operator. The total
hamiltonian of the system is given by Eq.(\ref{21}), the free
Hamiltonian of the scalar field and the Eq.(\ref{33}), where
$\lambda$ is a small coupling constant between the qubit and the
quantized Bose field.

We would like to stress that in general, measurement and state
preparation are different phenomena. In quantum mechanics
preparing a particular state might involve a special type of
measurement, but there are an infinite number of prepared states
which are not associated with measurements. For example, a generic
state for the two-level system $|\,\Psi\rangle$ can be written as
$|\,\Psi\rangle=\alpha\,|\,e\rangle +\beta |\,e\rangle$. The
normalization condition gives $|\,\alpha|^{2}+ |\,\beta|\,^{2}=1$.
We can also prepare the two-level system in another state which is
not an eigenstate of the Hamiltonian $H_{Q}$. Introducing the
variables $\theta$ and $\phi$ we can write
\begin{equation}
|\,\Psi\rangle=|\,\theta,\phi\rangle=e^{-i\frac{\phi}{2}}\cos\left(\frac{\theta}{2}\right)
|\,e\rangle + e^{i\frac{\phi}{2}}\sin\left(
\frac{\theta}{2}\right)|\,g\rangle.
\end{equation}
Clearly $|\,\theta,\phi\rangle$ is a complete set, since
$\int_{0}^{\pi}\,d\theta \sin
\theta\int_{0}^{2\pi}\frac{d\phi}{2\pi}\,
|\,\theta,\phi\rangle\,\langle\,\theta,\phi| = 1$. As already
stated, we are interested to study the probability of decay from
the excited state, driven by the vacuum fluctuations. Therefore we
assume that the two-level system is in an eigenstate of the
Hamiltonian $H_{Q}$. To evaluate the probability of decay
(excitation) of the two-level system interacting with the Bose
field, we can define the prepared initial state of the system in
$\tau=0$ as
$|\,\tau_{0}\,\rangle=|\,e\,\rangle\,\otimes\,|\,\Phi_{i}\,\rangle$
$(|\,\tau_{0}\,\rangle=|\,g\,\rangle\,\otimes\,|\,\Phi_{i}\,\rangle)$,
where $|\,\Phi_{i}\,\rangle$ is the initial state of the field.
Both situations can be analyzed using the same formalism. In the
interacting picture, using the first-order approximation the
probability of transition $P(E,\tau,0)$ after the time interval
$\tau$ is given by
\begin{equation}
P(E,\tau,0)=\lambda^{2}|\langle \, e \, | \, m(0) \, | \, g \,
\rangle|\,^{2}\,F(E,\tau,0),
\label{38}
\end{equation}
where the response function $F(E,\tau,0)$ is given by
\begin{equation}
F(E,\tau,0)=\int_{0}^{\tau}\,d\tau'\,\int_{0}^{\tau}\,d\tau''\,e^{-iE(\tau'-\tau'')}
\langle \, \Phi_{i}\,
|\,\varphi(x(\tau'))\,\varphi(x(\tau''))|\,\Phi_{i}\,\rangle,
 \label{39}
\end{equation}
and in the above equation $E=\pm\,\omega$, where the signs $(+)$
and $(-)$ represent the excitation and decay process,
respectively. Note that to obtain Eq.(\ref{39}) we sum over all
possible final states of the Bose field and we are using the
completeness relation $\sum_{f}|\,\Phi_{f}\,\rangle\,\langle \,
\Phi_{f}|=1$, where $|\,\Phi_{f}\,\rangle$ is an arbitrary Bose
field final state. We are obtaining information about the time
evolution of the sub-system, i.e., the two-level system. This
approach must be equivalent to study the density operator of the
sub-system, i.e., the two-level system, which describes the
dynamic evolution of this sub-system interacting with the
environment through master equations. The equivalence between the
density operator description and first-order perturbation theory
to evaluate the quantum non-decay probability was demonstrated a
long time ago by Fonda et al \cite{fonda}.

To proceed, let us suppose that it is possible to prepare the
scalar field in Poincar\'e invariant vacuum state
$|\,0,M\,\rangle$, or the Minkowski vacuum state. Therefore in the
above equation the quantity $\langle \, \Phi_{i}\,
|\,\varphi(x(\tau'))\,\varphi(x(\tau''))|\,\Phi_{i}\,\rangle$
becomes the positive Wightman function associated with the Bose
scalar field evaluated in the world-line of the qubit. There are
two points that we would like to stress. The first is that in the
integrand of Eq.(\ref{39}), the two-point correlation function
depends only on the time difference $(\tau'-\tau'')$. The
integration over $\tau'$ and $\tau''$ is carried out over the
square $0\leq\,\tau'\leq\,\tau$, $0\leq\,\tau'\leq\,\tau$
\cite{ss}. The second one is that we are not using the
rotating-wave approximation, used by Glauber \cite{glauber} and
others, to define an ideal photo-counter detector. Therefore in
the response function $F(E,\tau,0)$ the vacuum fluctuations
contributions associated with the Bose field are taken into
account and we are studying the radiative processes associated to
the qubit induced by a vacuum field.

To calculate the probability of transition evaluated in a finite
time interval let us prepare the system in the initial instant of
time $\tau_{i}$ in the state
\begin{equation}
|\,\tau_{i}\,\rangle=|\,e\,\rangle\,\otimes\,|\,0,M\,\rangle,
\label{40}
\end{equation}
and assume that we observe the system in the ground state in the
instant of time $\tau_{f}$. In this situation the response
function becomes
\begin{equation}
F^{(1)}(E,\tau_{f},\tau_{i})=\int_{\tau_{i}}^{\tau_{f}}\,d\tau'\,
\int_{\tau_{i}}^{\tau_{f}}\,d\tau''\,e^{-iE(\tau'-\tau'')} \langle
\, 0,M \, |\,\varphi(x(\tau'))\,\varphi(x(\tau''))|\,0,M\,\rangle,
 \label{41}
\end{equation}
where $E=-\omega$. We are using the subscript $(1)$ to call the
attention that this quantity is evaluated in the first
measurement. Defining $\Delta\tau=\tau_{f}-\tau_{i}$, and
introducing the variables $\xi = \tau' - \tau''$ and $\eta = \tau'
+ \tau''$, the response function given by Eq.(\ref{41}) can be
written as:
\begin{equation}
F^{(1)}(E,\Delta \tau) = - \frac{1}{4\pi^2}\int_{-\Delta
\tau}^{\Delta \tau} d\xi (\Delta \tau - | \, \xi| \, )
\frac{e^{-iE\xi}}{(\xi - i\varepsilon)^2},
 \label{42}
\end{equation}
where the $i\varepsilon$ is introduced to specify correctly the
singularities of the Wightman function, to respect causality
requirements.
Let us split the response function $F^{(1)}(E,\Delta \tau)$ in two
contributions:
\begin{equation}
F^{(1)}(E,\Delta \tau) = F^{(1)}_{1}(E,\Delta \tau) +F^{(1)}_{2}
(E,\Delta \tau), \label{43}
\end{equation}
where the functions $F^{(1)}_{1}(E,\Delta \tau)$ and
$F^{(1)}_{2}(E,\Delta \tau)$ are given respectively by
\begin{equation}
F^{(1)}_{1}(E,\Delta \tau) = - \frac{1}{4\pi^2}\int_{-\Delta
\tau}^{\Delta \tau} d\xi \Delta \tau \frac{e^{-iE\xi}}{(\xi -
i\varepsilon)^2} \label{44}
\end{equation}
and
\begin{equation}
F^{(1)}_{2}(E,\Delta \tau) = \frac{1}{4\pi^2}\int_{-\Delta
\tau}^{\Delta \tau} d\xi \, | \, \xi | \frac{e^{-iE\xi}}{(\xi -
i\varepsilon)^2}. \label{45}
\end{equation}
After some calculations \cite{ss} \cite{errata} we obtain that the
functions $F^{(1)}_{1}(E,\Delta \tau)$ and $F^{(1)}_{2}(E,\Delta
\tau)$ can be written as:
\begin{equation}
F^{(1)}_{1}(E,\Delta \tau) = \frac{\Delta \tau}{2\pi} \left( -E \,
\Theta (-E) + \frac{\cos E \Delta \tau}{\pi \Delta \tau} +
\frac{|E|}{\pi} \left(Si |E| \Delta \tau - \frac{\pi}{2} \right)
\right) \label{46}
\end{equation}
and
\begin{equation}
F^{(1)}_{2}(E,\Delta \tau) = \frac{1}{2\pi^2} \left( -\gamma +
Ci|E|\Delta \tau - \ln\,\varepsilon |E| - 1 \right). \label{47}
\end{equation}
In the Eq.(\ref{46}) and Eq.(\ref{47}), $\gamma$ is the Euler
constant and the $Si(z)$ and $Ci(z)$ functions are defined
respectively by \cite{stegun}:
\begin{equation}
Si(z) = \int_{0}^{z} \frac{\sin \, t}{t} \, dt,
 \label{48}
\end{equation}
and
\begin{equation}
Ci(z) = \gamma + \ln z + \int_{0}^{z} \frac{1}{t}\,(\cos t - 1) \,
dt. \label{49}
\end{equation}
The Eq.(\ref{47}) has two divergences. One given by $\ln \Delta
\tau$ as $\Delta \tau \rightarrow 0^{+}$ and other given by $\ln
\varepsilon$. In a full perturbative renormalizable quantum field
theory, there is a regularization and also a renormalization
procedure, where the infinities can be eliminated. One way to
circumvent this problem is to define the rate $R^{(1)}(E,\Delta
\tau) = \frac{d}{d (\Delta \tau)} F^{(1)}(E, \Delta \tau)$
\cite{string}. Since we are interested in the non-decay
probability in a finite time, let us define the renormalized
probability of transition $P^{(1)}_{ren}(E,\Delta \tau)$ by
\begin{equation}
P^{(1)}_{ren}(E,\Delta \tau)=\lambda^{2}|\langle \, e \, | \,
m(\tau_{i}) \, | \, g \, \rangle|\,^{2}\,\left( F^{(1)}(E,\Delta
\tau) - \ln \frac{\Delta \tau}{\varepsilon} \right). \label{50}
\end{equation}
To support this procedure we can use the argument that these
divergences are spurious and can not appear in the physically
measured processes. Using this renormalization procedure, the
probability of decay can be written as
\begin{equation}
P^{(1)}_{ren}(E,\Delta \tau) = \lambda^{2}|\langle \, e \, | \,
m(\tau_{i}) \, | \, g \, \rangle|\,^{2}\, F^{(1)}_{ren}(E,\Delta
\tau), \label{51}
\end{equation}
where $F^{(1)}_{ren}(E,\Delta \tau)$ is given by
\begin{equation}
F^{(1)}_{ren}(E,\Delta \tau) = \frac{1}{2\pi^2} \left( |E| \Delta
\tau \left( \frac{\pi}{2} + Si \, |E| \, \Delta \tau \right) +
\cos \, E \Delta \tau  -1 + \int_{0}^{|E|\,\Delta \tau}
\frac{1}{\xi}\,( \cos \xi -1)\,d\xi \right). \label{52}
\end{equation}
For a small time interval $\Delta \tau$ the transition probability
contains two contributions: the first one that increases linearly
with the time interval and the second one that increases
quadratically with the time interval. The probability of non-decay
of this two-level system after a finite time interval $\Delta
\tau$ is given by
$P^{(1)}_{still}(E,\Delta\tau)=(1-P^{(1)}_{ren}(E,\Delta \tau))$.

Instead of using the interaction picture and the perturbation
theory in first-order approximations, it is possible to use the
Heisenberg equations of motion to the Dicke operators and also to
the annihilation and creation operators associated to the Bose
field \cite{roc1} \cite{roc2}. Clearly both methods of
calculations must give identical results.

In the next Section we use this probability of non-decay to show
that the observed probability is exponential of the time $T$.
There are some technical problems in the second measurement.
Although we assume that the interaction between the qubit and the
field is weak, we can not suppose that the state of the field does
not change in time. We conclude that to study the time evolution
of the system in the second measuremet, we have to assume that the
qubit is still in the excited state $| \, e \, \rangle$, and the
Bose field is in an arbitrary state $|\,\Phi_{i}\rangle$. We will
assume for simplicity that the initial sate of the field in the
second measurement is a many particle state.

\section{The exponential decay after $N$ successive measurements }\

The aim of this section is to show that if we couple the two-level
system with the Bose field in the vacuum state we recover the
exponential non-decay probability if the system is continuously
observed.

Suppose that we perform $N$ measurements at equal time interval
which satisfies $T = N \Delta\tau$, and in each measurement the
system stays in the initial excited state $|\, e \, \rangle$
defined before. Again, we would like to stress that to obtain the
Eq.({\ref{39}) we assumed that no observation was made to
discriminate among possible final Bose field states. Therefore to
calculate the probability of observing the initial state of the
qubit $| \, e \, \rangle$, at the final time $T$ after $N$
measurements we have to point out the following fact. In the first
measurement, to obtain the probability
$P^{(1)}_{still}(E,\Delta\tau)=(1-P^{(1)}_{ren}(E,\Delta\tau))$,
where $P^{(1)}_{ren}(E,\Delta\tau)$ is given by Eq.(\ref{51}) and
Eq.(\ref{52}), we summed over all possible final states of the
Bose field. This reflect the fact that we are interested in the
final state of the qubit and not that of the field. For this
reason, in the second measurement we can not assume that the
initial state of the Bose field is identical with the initial
state of the field prepared in the first time interval
$\Delta\tau$.

Before the second measurement, the information of the state of the
field is retained by the system. Since the initial Bose field
state in the second measurement is indeterminate for us, we choose
an arbitrary state $|\,\Phi_{i} \, \rangle$. Therefore to study
the time evolution of the system in the second measuremet, let us
suppose that the qubit is still in the excited state $| \, e \,
\rangle$, and the Bose field is in an arbitrary state $|\,\Phi_{i}
\, \rangle$. To find the probability of transition from the state
$|\,e\,\rangle\,\otimes\,|\,\Phi_{i}\,\rangle$  to the final state
$|\,g\,\rangle\,\otimes\,|\,\Phi_{f}\,\rangle$ after the second
time interval $\Delta\tau$ we have to evaluate the expression
$P^{(2)}(E,\tau_{f}+\Delta\tau,\tau_{i}+\Delta\tau)$ given by
\begin{equation}
P^{(2)}(E,\tau_{f}+\Delta\tau,\tau_{i}+\Delta\tau)=\lambda^{2}|\langle
\, e \, | \, m(\tau_{i}+\Delta\tau) \, | \, g \,
\rangle|\,^{2}\,F^{(2)}(E,\tau_{f}+\Delta\tau,\tau_{i}+\Delta\tau),
\label{aaa1}
\end{equation}
where the new response function
$F^{(2)}(E,\tau_{f}+\Delta\tau,\tau_{i}+\Delta\tau)=F^{(2)}(E,\Delta\tau)$
can be written as
\begin{equation}
F^{(2)}(E,\Delta\tau)=\int_{-\Delta \tau}^{\Delta \tau} d\xi
(\Delta \tau - | \, \xi| \, ) e^{-iE\xi}
 \langle\, \Phi_{i}\,
|\,\varphi(x(\tau'))\,\varphi(x(\tau''))|\,\Phi_{i}\,\rangle.
 \label{aaa2}
\end{equation}
Note that we are using the same convention used in the section IV.
We have again that $\Delta\tau=\tau_{f}-\tau_{i}$, and $\xi =
\tau' - \tau''$. Note that to obtain the Eq.(\ref{aaa2}) we are
using again the completeness relation over the final states of the
field in the second measurement.

We have a considerable arbitrariness in the choice of the initial
state of the field after the first measurement. An appropriate
starting point is to suppose that the initial state of the field
in the second measurement is a many particle state with $n_{1}$
quanta with momenta $\bf{k}_{1}$ and energy $\omega_{1}$, $n_{2}$
quanta with momenta $\bf{k}_{2}$ and energy $\omega_{2}$ and so
on. Therefore the two-point correlation function that appears in
the Eq.(\ref{aaa2}) can be written as
\begin{equation}
\langle\, \Phi_{i}\,
|\,\varphi(x(\tau'))\,\varphi(x(\tau''))|\,\Phi_{i}\,\rangle=
\langle
n_{1}({\bf{k}}_{1})...n_{j}({\bf{k}}_{j})|\,\varphi(x(\tau'))\,\varphi(x(\tau''))|
\,n_{1}({\bf{k}}_{1})...n_{j}({\bf{k}}_{j})\rangle.
 \label{aaa3}
\end{equation}
Using Eq.(\ref{aaa3}) it's not difficult to show that we can write
the two-point correlation function $\langle\, \Phi_{i}\,
|\,\varphi(x(\tau'))\,\varphi(x(\tau''))|\,\Phi_{i}\,\rangle$ in
the following way:
\begin{equation}
G^{+}(x(\tau '),x(\tau '')) + \sum_i n_i u_{{\bf{k}}_i}(x(\tau '))
u^{*}_{{\bf{k}}_i}(x(\tau '')) + \sum_i n_i
u^{*}_{{\bf{k}}_i}(x(\tau ')) u_{{\bf{k}}_i}(x(\tau '')) ,
\label{aaa4}
\end{equation}
where $G^{+}(x(\tau '),x(\tau ''))$ is the positive Wightman
function evaluated in the world line of the qubit and $n_i$ is the
number density of quanta in the $k$-space. The set
$({u^{*}_{{\bf{k}}_i}(x), u_{{\bf{k}}_i}(x)})$ is a basis in the
space of solutions of the Klein-Gordon equation. Without loss of
generality we can choose the plane-waves for the basis.

Taking the continuous limit, assuming that the quanta are
distributed isotropically, and substituting Eq.(\ref{aaa3}) and
Eq.(\ref{aaa4}) in Eq.(\ref{aaa2}) we have that the response
function $F^{(2)}(E, \Delta \tau)$ in the second measurement can
be written as
\begin{equation}
F^{(2)}(E, \Delta \tau) = F^{(2)}_{1}(E, \Delta \tau) +
F^{(2)}_{2}(E, \Delta \tau). \label{aaa5}
\end{equation}
In the Eq.(\ref{aaa5}), the quantity $F^{(2)}_{1}(E, \Delta \tau)$
is the vacuum contribution given by Eq.(\ref{42}) and the quantity
$F^{(2)}_2(E, \Delta \tau)$ is the non-vacuum contribution given
by
\begin{equation}
F^{(2)}_{2}(E,\Delta \tau) = \frac{1}{4 \pi^2}\int_{-\Delta
\tau}^{\Delta \tau} d\xi \, (\Delta \tau - | \, \xi| \, )
e^{-iE\xi} \, g(\xi). \label{aaa6}
\end{equation}
In the above expression the function $g(\xi)$ that depends on the
the number density of quanta in the $k$-space in the continuous
limit is
\begin{equation}
g(\xi) = \int d\omega \, \omega \, n(\omega) \left( e^{i \omega
\xi} + e^{-i \omega \xi} \right).
\label{aaa7}
\end{equation}
To proceed, we can extend the integration over all frequencies in
the Eq.(\ref{aaa7}) and also replace the number density of quanta
in the $k$-space, $n(\omega)$ by a constant value in the interval
$[0,a]$. Using that \cite{tabela}
\begin{equation}
\int^{\infty}_{-\infty} dx \, f_n(x) \, e^{ix \xi} =
\frac{n!}{(-i\xi)^{n+1}} \left( 1 + e^{ia\xi} \sum^n_{k=0}
\frac{(-ia\xi)^k}{k!} \right),
 \label{aaa8}
\end{equation}
where $f_n(x) = x^n$ for $0<x<a$ and zero otherwise and $n=
1,2,...$ we get that Eq.(\ref{aaa7}) can be written as
\begin{equation}
g(\xi,a) = -\frac{2}{\xi^2} -\frac{1}{\xi^2} \left( e^{ia\xi} +
e^{-ia\xi} \right) - \frac{2a}{\xi} \, \sin{\xi a}. \label{aaa9}
\end{equation}
The first term in the above equation will give a contribution to
the response function which is proportional to the one of the
vacuum field. To proceed, let us substitute Eq.(\ref{aaa9}) in
Eq.(\ref{aaa6}). Therefore we have that $F^{(2)}_{2}(E, \Delta
\tau)$ can be written as
\begin{equation}
F^{(2)}_{2}(E,a, \Delta \tau) =  f^{(2)}_{1}(E,\Delta \tau) +
f^{(2)}_{2}(E,a, \Delta \tau) + f^{(2)}_{3}(E,a, \Delta \tau) +
f^{(2)}_{4}(E,a, \Delta \tau)), \label{aaa10}
\end{equation}
where $f^{(2)}_1(E, \Delta \tau) = 2 F^{(2)}_{1}(E, \Delta \tau)$,
and the two other terms in the above equation are given
respectively by
\begin{equation}
f^{(2)}_{2}(E,a, \Delta \tau) = -\frac{1}{4\pi^2} \int^{\Delta
\tau}_{-\Delta \tau} d\xi \, (\Delta \tau - |\, \xi \, |)
\frac{e^{-i(E+a)\xi}}{(\xi - i\epsilon)^2}, \label{aaa11}
\end{equation}
and $f^{(2)}_{3}(E,a, \Delta \tau)= f^{(2)}_{2}(E,-a, \Delta
\tau)$.
Again the $i\varepsilon$ is introduced to respect causality
requirements. Finally, the function $f^{(2)}_{4}(E,a, \Delta
\tau)$ is given by
\begin{equation}
f^{(2)}_{4}(E,a, \Delta \tau) = -\frac{a}{4\pi^2}\int^{\Delta
\tau}_{-\Delta \tau} d\xi (\Delta \tau - | \, \xi \, |) \,
e^{-iE\xi} \, \frac{\sin{\xi a}}{(\xi - i\epsilon)}. \label{aaa13}
\end{equation}
The integral above can be written in the following form:
\begin{equation}
f^{(2)}_{4}(E,a, \Delta \tau)=-\frac{a\Delta\tau}{8\pi
i}\left(\int^{\Delta \tau}_{-\Delta \tau} \frac{d\xi}{\xi} \,
{e^{-i\xi(E-a)}}- \int^{\Delta \tau}_{-\Delta
\tau}\frac{d\xi}{\xi}\,e^{-i\xi(E+a)}
\right)+\frac{a}{2\pi}\int_{0}^{\Delta\tau}d\xi \sin\xi a\cos\xi
E. \label{aaa14}
\end{equation}
To carry out the integrations we use the fact that the two
integral in the right side of the Eq.(\ref{aaa14}) can be
interpreted as the principal value, i.e.
\begin{equation}
\epsilon (x) =\frac{1}{i\pi }P\int^{\infty
\tau}_{-\infty} \frac{d\xi}{\xi} \, {e^{i\xi
x}}
\label{aaa15}
\end{equation}
and also the last integral of the Eq.(\ref{aaa14}) can be carry
out immediately and gives
\begin{equation}
\int_{0}^{\Delta\tau}d\xi \sin\xi a\cos\xi
E=\frac{1}{2(E-a)}\biggl(\cos(E-a)\Delta\tau-1\biggr)-
\frac{1}{2(E+a)}\biggl(\cos(E+a)\Delta\tau-1 \biggr).
\label{aaa16}
\end{equation}
We conclude that in this second small time interval $\Delta \tau$
the transition probability contains also two contributions: the
first one that increases linearly with the time interval and the
second one that increases quadratically with the time interval.
Following this line, the probability of observing the initial
state at the final time $T$ after $N$ measurements reads
\begin{equation}
P_{still}^{(N)}(E,T) = \left[ P^{(1)}_{still} \left(E, \frac{T}{N}
\right) \right]\left[P^{(2)}_{still}(E,a,
\frac{T}{N})\right]^{(N-1)}. \label{aqdecay3}
\end{equation}
Using the fact that
$P^{(i)}_{still}\left(E,\left(\frac{T}{N}\right)\right)=\left
[1-P^{(i)}_{ren}\left(E,\left(\frac{T}{N}\right)
\right)\right]$ for $i=1,2$, we have
\begin{equation}
P_{still}^{(N)}(E,T) =
\left[1-P^{(1)}_{ren}\left(E,\left(\frac{T}{N}\right)
\right)\right] \left[1 - P^{(2)}_{ren}(E,a,
\frac{T}{N})\right]^{(N-1)}. \label{aqdecay4}
\end{equation}
For $\Delta \tau = \frac{T}{N}$, expanding for small arguments and
keeping terms only through order $\left(\frac{T}{N}\right)^{2}$
the quantity $F^{(1)}_{ren}(E,\Delta \tau)$ is written as
\begin{equation}
F^{(1)}_{ren}\left(E,\frac{T}{N}\right) \approx \left(
\frac{|E|T}{4\pi N} + \frac{\alpha E^2 T^2}{4\pi N^2} \right),
\end{equation}
where $\alpha = (\frac{1}{2} - \frac{3}{2\pi})$. The quantity
$F^{(2)}_{ren}(E,\Delta \tau)$ also in the same order of the
approximation can be written as
\begin{equation}
F^{(2)}_{ren}\left(E,a,\frac{T}{N}\right) \approx
\left(\,p\,(E,a)\frac{T}{N} + q(E,a)\frac{T^{2}}{N^{2}} \right),
\end{equation}
where $p\,(E,a)$ and $q\,(E,a)$ are functions of $E$ and $a$ given
by
\begin{equation}
p\,(E,a) = \frac{1}{4\pi} \biggl( 3|E| + |E +a| + |E-a| +a\,
\theta (a-E) - a\,\theta (-E-a) + a\, \theta (E+a) \biggr)
\end{equation}
and
\begin{equation}
q\,(E,a) = \frac{1}{8\pi^2} \left( 5E^2(\pi -3) - 2a^2\right).
\end{equation}
It is easy to see that, if $|E|<|\,a|$, then
\begin{equation}
p\,(E,a) = \frac{1}{4\pi} \left( 3|E| + 4a \right).
\end{equation}
For $|E|>|\,a|$:
\begin{equation}
p\,(E,a) = \frac{5|E|}{4\pi}.
\end{equation}
Finally for $|E|=|\,a|$, we obtain
\begin{equation}
p\,(E,a) = \frac{|E|}{\pi}
\end{equation}
and
\begin{equation}
q\,(E,a) = (5\pi-17)\frac{E^2}{8\pi^2}.
\end{equation}
Defining $\lambda^{2}|\langle \, e \, | \, m(\tau_{i}) \, | \,
g \, \rangle|\,^{2} = \sigma$, we may write the probability of
observation the initial state at a finite time $T$ after $N$
measurements $P_{still}^{(N)} (E,\frac{T}{N})$ as
\begin{equation}
P_{still}^{(N)} (E,a,N) \approx \left[ 1 - \frac{1}{N} \left(
p\,(E,a,\sigma)T + q(E,a,\sigma)\frac{T^{2}}{N} \right) \right]^N.
\label{aaa}
\end{equation}

At this moment we would like to discuss the interpretation of
time-uncertainty relations. There are a large amount of literature
devoted to the interpretation of quantum mechanics. Nevertheless,
concerning the time-uncertainty relations, there are a few papers
discussing the implications of such relations. Landau and Peierls
\cite{landau1} and also Landau and Lifshitz \cite{landau2} claim
that the energy of a quantum system can be measured exactly at a
given time. Nevertheless, we must take into account the change
caused by the process of measurement. In the relation $\Delta
E\,\Delta \tau>1$, the quantity $\Delta E$ is the difference
between two exactly measured energy values at two different
instants of time, where $\Delta \tau$ is the time interval between
the measurements. If we accept this interpretation, there is a
finite but very large N constrained by an upper bound given by the
Landau, Peierls and Lifshitz interpretation of the time-energy
uncertainty relation $(N<TE)$, and we obtain that the non-decay
probability is polinomial but very similar to the exponential
behavior.

On the other hand, this interpretation of the time-energy
uncertainty relation is not universally accepted. Aharonov and
Bohm \cite{aharonov} \cite{bohm} claim that the time-uncertainty
relations are not consistent with the general principles of
quantum mechanics which require that the uncertainty relations be
expressible in terms of operators. Therefore, they concluded that
the energy of a quantum system can be measured in an arbitrary
short time. In this framework we are able to take the limit $N
\rightarrow \infty$, and we get
\begin{equation}
P_{still}^{(N)}(E,a,T) = \, \exp{\left(-p\,(E,a,\sigma)T -
q(E,a,\sigma)\frac{T^{2}}{N}\right)}. \label{aqdecay5}
\end{equation}
To conclude, we obtain that repeated observations slow-down the
evolution of the unstable system and increase the probability that
the system remains in the initial state at $T$. If we are able to
perform only a finite but very large N number of measurements in a
finite time, constrained by an upper bound given by the Landau,
Peierls and Lifshitz interpretation of the time-energy uncertainty
relation, we obtain that the non-decay probability is polinomial
but very similar to the exponential behavior. On the other hand,
if we are able to perform continuous observation and the limit
$N\rightarrow \infty$ can be used, the non-decay probability
becomes a pure exponential.

To obtain this classical behavior, we follow different steps.
First, we prepare the composed system in an initial state
$|\,e\,\rangle\,\otimes\,|\,0,M\,\rangle$, i.e., we couple the
two-level system with a vacuum field. Second, that the system
evolves under the influence of the unobserved Bose field. Finally
that we are able to perform continuous observations in the
two-level system.

\section{Conclusions}

In this paper we study the time evolution of unstable systems
after repeated but finite observations and also in the limit of
continuous observation. We show that a continuously observed
quantum system has a classical time evolution behavior if it
interacts with a unobserved vacuum field.

Using perturbation theory in first-order approximation where a
two-level system is interacting with the Bose field in the
Poincar\'e invariant vacuum state we obtain two distinct types of
behaviors. First a finite but very large $N$ constrained by an
upper bound given by the Landau, Peierls and Lifshitz
interpretation of the time-energy uncertainty relation, and
second, the case where $N \rightarrow \infty$, allowed by the
Aharanov-Bohm interpretation of the same relation. Studying the
non-decay probability in both situations, we obtain that the
non-decay probability is polynomial and very similar to the
exponential behavior for the first case. For the second case the
non-decay probability is given by
\begin{equation}
P^{(N)}_{still}(E,T)=\exp{\left(-\frac{T}{\tau_{c}}\right)},
\end{equation}
where $\tau_{c} \propto \frac{1}{E}$. It is important to remind
that in the interaction Hamiltonian for our model, we are not
assuming the rotating-wave-approximation, that excludes terms that
represent simultaneous qubit and field excitation and
de-excitation respectively. Although these terms are not-energy
conserving, representing virtual processes, we understand that
only for very large time interval $\Delta\tau$ the contribution
coming from these terms can be neglected. In the case that we are
interested, i.e., a small $\Delta\tau$, a more carefully procedure
is not to omit the energy-non-conserving terms. This procedure
allow us to obtain the above equation that accounts very well for
experimental observed facts, as for example the decay of many
quantum systems, as unstable atoms or nuclei. The result
establishes that the quantum theory allow us to recover classical
behavior under suitable circumstances.

From the preceding sections it is seen that, although our model of
the qubit-Bose field composed system is quite satisfactory to
obtain the experimentally observed non-decay probability law with
the exponential behavior, the approach of the paper is
intrinsically limited since we would like to predict the same
behavior in quite general quantum systems. We know that many
systems have a complete set of discrete eigenstates but also a
continuum spectrum. Therefore to construct a more realistic model
to study radiative processes of unstable systems we have to
generalize our model to one with two bound states and also a
continuum of states. For example in the case of the atom, which is
a practical photo-detector, there is a continuum of final electron
states. Assuming the same two-levels and a continuum of states
$|\,\omega_{a}\rangle$, with energy in the range
$[\,\omega_{c},\infty\,]$, and preparing the small system in the
state$ |\,e\rangle$ and again the Bose field in the vacuum state
we have that the probability, evaluated for a finite time
interval, of the system to makes a transition to the continuum is
given  by
\begin{equation}
P(\Delta\tau)=\int_{\omega_{c}}^{\infty}\,d\\\omega_{a}\,\rho(\omega_{a})P(\omega_{ae},\Delta\tau),
\label{cont1}
\end{equation}
where $\omega_{ae}=\omega_{a}-\omega_{e}$ and $\rho(\omega_{a})$
is a density of final excited states. Again, the quantity
$P(\omega_{ae},\Delta\tau)$ is given by
\begin{equation}
P_{ren}(\omega_{ae},\Delta \tau) = \lambda^{2}|\langle \, a \, |
\, m(\tau_{i}) \, | \, e \, \rangle|\,^{2}\,
F_{ren}(\omega_{ae},\Delta \tau).
\label{cont2}
\end{equation}
Note that we have to choose a particular form to the density of
final excited states $\rho(\omega_{ae})$, to make sure that the
integral given by Eq.(\ref{cont1}) converges at infinity. This
generalization is under investigation by the authors.

There are also different directions for investigation. To mention
a few: first is to assume a strong-coupling between the qubit and
the Bose field \cite{strong1} \cite{strong2}. Second, still in the
weak-coupling regime, is to assume that the reservoir is in
thermal equilibrium or in a squeezed state \cite{stephany}. It can
be shown that the behavior of the two-level system in a squeezed
bath depends on the way in which the squeezed bath is prepared,
showing Zeno or anti-Zeno effects. Also it is interesting to
consider $N$ qubits $(N \rightarrow \infty)$ interacting with one
mode of the Bose field, analyzing the situation where the qubit
system acts as a reservoir whereas the Bose field is an open
system, and study the dynamics of the reduced system.

Finally, this two-level system, referred to as a qubit is the
elementary building block of a quantum computer \cite{albert}
\cite{fey1} \cite{fey2} \cite{deutsch} \cite{steane}. This new
area of research has revived the interest in open quantum systems.
The fundamental technological problem is if it is possible to
create entanglement properties of states in systems that interact
with a reservoir. Several situations of entangled systems have
been proposed, as for example involving trapping and also cooling
a small number of atoms. How to isolate atoms from the environment
in order to make the effect of decoherence negligible is an open
problem until now.

Therefore, another natural extension of this paper is to
generalize some results of the paper in the case of two-atom
systems prepared in an entangled state \cite{tanas}. Using
time-dependent perturbation theory in a first-order approximation,
evaluate the probability per unit-time of decay of the symmetric
and anti-symmetric states given by Eq.(\ref{dipole2}) and
Eq.(\ref{dipole3}) respectively to the ground state $| \, g_{1} \,
\rangle\,\otimes\,| \, g_{2} \, \rangle$. Note that we have to
continue to assume that the qubits also interact with the vacuum
modes. The possibility to prepare the two-atom system in a
entangled decoherence-free state is a question that has
fundamental importance in quantum computing applications. Although
the spontaneous emission from a pair of atoms \cite{par1}
\cite{par2} and the causal aspects of their spontaneous decay
\cite{par3} \cite{par4} have been studied by some authors, the
extension of this formalism for two-atom prepared in an entangled
state, evaluating the probability of decay in a finite time
interval is new in the literature.

\section{Acknowlegements}

We would like to thank  S. Joffily, H. J. Mosqueira Cuesta and G.
Flores-Hidalgo for enlightening discussions. This paper was
supported by Conselho Nacional de Desenvolvimento Cientifico e
Tecnol{\'o}gico do Brazil (CNPq).

\begin{appendix}
\makeatletter \@addtoreset{equation}{section} \makeatother
\renewcommand{\theequation}{\thesection.\arabic{equation}}

\section{The qubit-Bose field interaction Hamiltonians}\

In this appendix we consider a very general situation where the
system under investigation contains a large number of
non-identical two-level systems. In order to describe the dynamics
of the reservoir and the two-level systems we have to introduce
the Hamiltonian governing the interaction of the quantized Bose
field with free qubits. Free means that there is no interaction
between the qubits. Therefore let us consider a Bose quantum
system $B$, with Hilbert space ${\cal H}^{(B)}$ which is coupled
with $N$ qubits, with Hilbert space  ${\cal H}^{(Q)}$. Let us
assume that the reservoir is in thermal equilibrium at temperature
$\beta^{-1}$. The Bose quantum system is a sub-system of the total
system living in the tensor product space ${\cal
H}^{(B)}\,\otimes\,{\cal H}^{(Q)}$.

Let us denote by $H_{B}$ the Hamiltonian of the quantized Bose
field, by $H_{Q}$  the free Hamiltonian of the $N$-qubits and
$H_{I}$ the Hamiltonian describing the interaction between the
quantized Bose field and the $N$ qubits. The Hamiltonian for the
total system can be written as
\begin{equation}
H=H_{B}\,\otimes\,I_{Q}+I_{B}\,\otimes\,H_{Q}+H_{I}, \label{13}
\end{equation}
where $I_{B}$ and $I_{Q}$ denotes the identities in the Hilbert
spaces of the quantized Bose field and the $N$ qubits.

The main purpose of this appendix is to discuss qubit-Bose field
interaction Hamiltonians. Therefore, let us introduce the Dicke
operators to describe each qubit. The free $j-th$ qubit
Hamiltonian will be denoted by $H_{D}^{(j)}$, since we are using
the Dicke representation. Therefore, we have
\begin{equation}
H_{D}^{(j)}|\,i\,\rangle_{j}=\omega_{i}^{(j)}|\,i\,\rangle_{j},
\label{14}
\end{equation}
where $|\,i\,\rangle_{j}$ are orthogonal energy eigenstates
accessible to the $j-th$ qubit and $\omega_{i}^{(j)}$ are the
respective eigenfrequencies. Using Eq.(\ref{14}) and the
orthonormality of the energy eigenstates we can write the  $j-th$
qubit Hamiltonian $H_{D}^{(j)}$ as
\begin{equation}
H_{D}^{(j)}=\sum_{i=1}^{2}\,\omega_{i}^{(j)} ( |\,i\,\rangle \,
\langle\, i| \, )_j . \label{15}
\end{equation}
Let us define the Dicke operators $\sigma_{(j)}^z$,
$\sigma_{(j)}^{+}$ and $\sigma_{(j)}^{-}$ for each qubit by
\begin{equation}
\sigma_{(j)}^z=\frac{1}{2} \, ( |2\,\rangle \, \langle\,2|-
\,|1\,\rangle \, \langle\,1| \, ) _j , \label{16}
\end{equation}
\begin{equation}
\sigma_{(j)}^{+}=\,(|2\,\rangle \, \langle\,1| \, )_j , \label{17}
\end{equation}
and finally
\begin{equation}
\sigma_{(j)}^{-}=\, (|1\,\rangle \, \langle\,2| \, )_j .
\label{18}
\end{equation}
The Dicke representation is a second quantization of the qubits.
Combining Eq.(\ref{15}) and Eq.(\ref{16}), the $j-th$ qubit
Hamiltonian can be written as
\begin{equation}
H_{D}^{(j)}=\,\Omega^{(j)}\,\sigma_{(j)}^z+\frac{1}{2}
\biggl(\omega_{1}^{(j)}+\omega_{2}^{(j)}\biggr), \label{19}
\end{equation}
where the energy gap between the energy eigenstates of the $j-th$
qubit is given by
\begin{equation}
\Omega^{(j)}=\omega_{2}^{(j)}-\omega_{1}^{(j)}. \label{20}
\end{equation}
Shifting the zero of energy to
$\frac{1}{2}(\omega_{1}^{(j)}+\omega_{2}^{(j)})$ for each qubit,
the $j-th$ qubit Hamiltonian given by Eq.(\ref{19}) can be
rewritten as
\begin{equation}
H_{D}^{(j)}=\Omega^{(j)}\,\sigma_{(j)}^z. \label{21}
\end{equation}
Note that the operators $\sigma_{(j)}^{+}$, $\sigma_{(j)}^{-}$ and
$\sigma_{(j)}^z$ satisfy the standard angular momentum commutation
relations corresponding to spin $\frac{1}{2}$ operators, i.e.,
\begin{equation}
\left[\sigma_{(j)}^{+},\sigma_{(j)}^{-}\right]=2\,\sigma_{(j)}^z,
\label{22}
\end{equation}
\begin{equation}
\left[\sigma_{(j)}^z,\sigma_{(j)}^{+}\right]=\sigma_{(j)}^{+},
\label{23}
\end{equation}
and finally
\begin{equation}
\left[\sigma_{(j)}^z,\sigma_{(j)}^{-}\right]=-\sigma_{(j)}^{-}.
\label{24}
\end{equation}
A well-known model is a combining system where we have only one
mode of the quantized field. The Hamiltonian of the $j-th$ qubit
$H_{D}^{(j)}$, with the contribution of the one-mode quantized
Bose field $H_{S}$, and the interaction Hamiltonian $H_{I}^{(j)}$,
can be used to define the Hamiltonian of the total system, given
by
\begin{eqnarray}
&& I_B\,\otimes\, H_{D}^{(j)}+H_{B}\,\otimes\,
I_Q +H_{I}^{(j)}=\nonumber\\
&&
I_B\,\otimes\,\Omega^{(j)}\,\sigma_{(j)}^z+\omega_{0}\,a^{\dagger}a\,\otimes\,
I_Q+g\,\Bigl(a\, +\,
a^{\dagger}\Bigr)\otimes\,\Bigl(\sigma_{(j)}^{+}+\sigma_{(j)}^{-}\Bigr),
\label{25}
\end{eqnarray}
where the second term in the Eq.(\ref{25}) has the contribution
from the quantized Bose field single mode Hamiltonian and the last
term is the interaction Hamiltonian of the $j-th$ qubit with the
one-mode quantized field. In the Eq.(\ref{25}) $g$ is a small
coupling constant between the qubit and the one mode quantized
Bose field. The generalization to $N$ qubits is described by
\begin{eqnarray}
&&I_B\,\otimes\, \sum_{j=1}^{N}\, H_{D}^{(j)}+
H_{B}\,\otimes\,I_Q+
\sum_{j=1}^{N}\,H_{I}^{(j)}=\nonumber\\
&& I_B\,\otimes\,\sum_{j=1}^{N}\,
\Omega^{(j)}\,\sigma_{(j)}^z+\omega_{0}\,a^{\dagger}a\,\otimes\,I_Q+
\Bigl(a+a^{\dagger}\Bigr)\otimes\,\frac{g}{\sqrt{N}}
\sum_{j=1}^{N}\, \Bigl(\sigma_{(j)}^{+}+\sigma_{(j)}^{-}\Bigr),
\label{26}
\end{eqnarray}
where the first summation in the right hand side
\begin{equation}
\sum_{j=1}^{N}\, \Omega^{(j)}\,\sigma_{(j)}^z=
\Omega^{(1)}\,\sigma_{(1)}^z\,\otimes{\bf{1}}\,\otimes...\otimes{\bf{1}}+...+
{\bf{1}}\,\otimes{\bf{1}}\otimes...\otimes{\bf{1}}\otimes\,\Omega^{(N)}\,\sigma_{(N)}^z,
\label{def}
\end{equation}
and $\bf{1}$ denotes the identity in the Hilbert space of each
qubit. We can also introduce a qubit-qubit interaction, which is
relevant in the study of entangled states. In an entangled system,
the state of the composite system can not be factorized in to a
product of the states of its sub-systems. For example in the case
of two-atom systems, takes the form
\begin{equation}
H_{(qq)}=\sum_{i\neq
j}^{2}H_{(ij)}\,\sigma_{(i)}^{+}\otimes\sigma_{(j)}^{-}.
\label{dipole}
\end{equation}
In the absence of the "dipole-dipole" interaction the pure Hilbert
space of the two qubit system is spanned by the states $| \, g_{1}
\, \rangle\,\otimes\,| \, g_{2} \, \rangle, | \, g_{1} \,
\rangle\,\otimes\,| \, e_{2} \, \rangle, | \, e_{1} \,
\rangle\,\otimes\,| \, g_{2} \, \rangle, | \, e_{1} \,
\rangle\,\otimes\,| \, e_{2} \, \rangle$, where $g$ and $e$
denotes respectively the ground and the excited state. If we
include the "dipole-dipole" interaction term in the form of
Eq.(\ref{dipole}), the vectors $| \, g_{1} \, \rangle\,\otimes\,|
\, e_{2} \, \rangle$ and  $| \, e_{1} \, \rangle\,\otimes\,| \,
g_{2} \, \rangle$ are not more eigenstates of the Hamiltonian of
the qubits systems. It can be shown that these two vectors states
must be substituted by the two entangled states, known in the
literature as maximally entangled states \cite{tanas}
\begin{equation}
| \, s \,\rangle=\frac{1}{\sqrt{2}}\,\left( | \, e_{1}
\,\rangle\,\otimes\,| \, g_{2} \, \rangle+ | \, g_{1}
\,\rangle\,\otimes\,| \, e_{2} \, \rangle\right)
 \label{dipole2}
\end{equation}
and
\begin{equation}
| \, a \,\rangle=\frac{1}{\sqrt{2}}\,\left( | \, e_{1}
\,\rangle\,\otimes\,| \, g_{2} \, \rangle- | \, g_{1}
\,\rangle\,\otimes\,| \, e_{2} \, \rangle\right).
 \label{dipole3}
\end{equation}
Going back to Eq.(\ref{26}), the interaction Hamiltonian is
simplified if we assume the Jaynes-Cummings model \cite{ja1}.
Considering the Jaynes-Cummings model for one qubit, we have
\begin{eqnarray}
&&
I_B\,\otimes\, H_{D}^{(j)}+H_{B}\,\otimes\,I_Q +H_{I}^{(j)}=\nonumber\\
&&
I_B\,\otimes\,\Omega^{(j)}\,\sigma_{(j)}^z+\omega_{0}\,a^{\dagger}a\,\otimes\,I_Q+
g\Bigl(a\,\otimes\,\sigma_{(j)}^{+}+a^{\dagger}\,\otimes\sigma_{(j)}^{-}\Bigr).
\label{27}
\end{eqnarray}
The generalization to $N$ qubits is straightforward and is given
by
\begin{eqnarray}
&& I_B\,\otimes\, \sum_{j=1}^{N}\,
H_{D}^{(j)}+H_{B}\,\otimes\,I_Q+\sum_{j=1}^{N}\,H_{I}^{(j)}=\nonumber\\
&&
I_B\,\otimes\,\sum_{j=1}^{N}\,\Omega^{(j)}\,\sigma_{(j)}^z+\omega_{0}\,a^{\dagger}a\,\otimes\,I_Q+
\frac{g}{\sqrt{N}}\sum_{j=1}^{N}\, \Bigl(a\,\otimes\,
\sigma_{(j)}^{+}\,+a^{\dagger}\,\otimes\,\sigma_{(j)}^{-}\Bigr).
\label{28}
\end{eqnarray}
One point which is important to stress is that the terms which we
ignore in Eq.(\ref{27}) and Eq.(\ref{28}) are the so called
counter-rotating terms. This approximation is known as the
rotating-wave-approximation. In the rotating-wave-approximation we
ignore energy non-conserving terms in which the emission
(absorption) of a quantum of a quantized field is accompanied by
the transition of one qubit from its lower (upper) to its upper
(lower) state.

So far we have discussed $N$ non-identical qubits interacting with
one-mode of the quantized Bose field. Our aim is now to discuss
the interaction of a system of $N$ identical qubits with energy
gap $(\Omega=\omega_{2}-\omega_{1})$, with an infinite number of
harmonic oscillators which defines the reservoir. Let
$b_{k}^{\dagger}$ and $b_{k}$ be the creation and annihilation
operators of the $k-th$ harmonic oscillator of frequency
$\omega_{k}$. The total Hamiltonian, i.e., the Hamiltonian of the
combined system of the reservoir and the $N$ identical qubits
interacting with the reservoir reads
\begin{equation}
I_B\,\otimes \Omega\sum_{j=1}^{N}\,\sigma_{(j)}^z +
\sum_{k} \omega_{k}\,b_k^{\dagger}\,b_k \,\otimes\,I_Q +
\frac{g}{\sqrt{N}}\sum_{j=1}^{N}\sum_{k} \Bigl(b_{k}\,\otimes\,\sigma_{(j)}^{+}
+b^{\dagger}_{k}\,\otimes\,\sigma_{(j)}^{-}\Bigr). \label{29}
\end{equation}
In the Eq.(\ref{29}) the first term is the free Hamiltonian of $N$
identical qubits, the second term is the free reservoir
Hamiltonian and the third term is the interaction Hamiltonian
between the reservoir and the $N$ identical qubits. Notice that we
shift the zero of energy for each qubits, as we did before, and we
are assuming the rotating-wave-approximation, where
$\frac{g}{\sqrt{N}}$ is the j-th qubit, k-th harmonic oscillator
coupling constant.

We can also use a different interaction Hamiltonian, as the one
introduced by Di Vicenzo \cite{di}. This author proposed a soluble
model to study the influence of decoherence in quantum computers,
with the following model describing a system of one qubit coupled
to a reservoir of harmonic oscillators:

\begin{eqnarray}
&&
I_B\,\otimes\,H_{Q}+H_{B}\,\otimes I_Q +H_{I}=\nonumber\\
&& I_B\,\otimes\,\Omega\,\sigma^z+\sum_{k}
\omega_{k}\,b_{k}^{\dagger}\,b_{k}\,\otimes I_Q+
g\sum_{k} \,\left(b_{k}^{\dagger}+ b_{k}\right)\otimes
\sigma^z, \label{30}
\end{eqnarray}
where $\Omega$ is the usual energy level spacing of the qubit,
$b_{k}^{\dagger}$ and $b_{k}$ are respectively the bosonic
creation and annihilation operators of the harmonic oscillators.
Notice the particular coupling between the reservoir and the
qubit, that allows the loss of quantum coherence induced by the
reservoir without affecting the qubit. There are two
straightforward generalizations for this model. The first one is
the introduction of a mode-dependent coupling constant \cite{hep}.
Therefore we have
\begin{eqnarray}
&&
I_B\,\otimes\, H_{Q} +H_{B}\,\otimes I_Q +H_{I}=\nonumber\\
&& I_B\,\otimes \Omega\,\sigma^z+
\sum_{k} \omega_{k}\,b_{k}^{\dagger}\,b_{k}\,\otimes I_Q
+ \sum_{k} \left(\lambda_{k}\,b_{k}^{\dagger}+
\lambda_{k}^{*}\,b_{k}\right)\,\otimes\sigma^z. \label{30}
\end{eqnarray}
Other straightforward generalization is to introduce $N$ identical
qubits and the Hamiltonian of the composed system reads
\begin{eqnarray}
&&
I_B\,\otimes\, H_{Q}+H_{B}\,\otimes I_Q+H_{I}=\nonumber\\
&&I_B\,\otimes\,\Omega\sum_{j=1}^{N}
\,\sigma_{(j)}^{z}+\sum_{k}
\omega_{k}\,b_{k}^{\dagger}\,b_{k}\,\otimes\,I_Q+
\frac{g}{\sqrt{N}}\sum_{j=1}^{N}
\sum_{k}\,\left(b_{k}^{\dagger}+ b_{k}\right)\,\otimes
\sigma_{(j)}^{z}. \label{31}
\end{eqnarray}
Another generalization of the Hamiltonian given by Eq.(\ref{29})
is not assume the rotating-wave-approximation in the interaction
Hamiltonian. Without the rotating-wave-approximation, the
interaction Hamiltonian between the $N$ qubits and the reservoir
of harmonic oscillators reads
\begin{equation}
H_{I}=\frac{g}{\sqrt{N}}\sum_{j=1}^{N}\sum_{k}
 \left(
b_{k}+b_{k}^{\dagger}\right)\,\otimes \left(\sigma_{(j)}^{+}+
\sigma_{(j)}^{-}\right). \label{32}
\end{equation}
To conclude this appendix we should point out that through the
paper we used the simple model where the interaction Hamiltonian
between the two-level system and the scalar field is linear in
both field and qubit and is given by
\begin{equation}
H_{I}=\lambda\left(m_{21}\,\sigma^{+}+m_{12}\,\sigma^{-}+\sigma^{z}(m_{22}-m_{11})
\right)\,\otimes\,\varphi(x), \label{33}
\end{equation}
where $m_{ij}=\langle \, i \, | \, m(0) \, | \, j \, \rangle$, and
$\lambda$ is a small coupling constant. The Bose field
$\varphi(x)$ can be expanded as
\begin{equation}
\varphi(x) = \sum_{\bf{k}} \left( a_{\bf{k}}
u_{\bf{k}}(t,{\bf{x}}) + a^{\dagger}_{\bf{k}}
u^{*}_{\bf{k}}(t,\bf{x}) \right).
\end{equation}
The modes $u_{\bf{k}}(t,{\bf{x}})$ form a basis in the space of
solutions of the Klein-Gordon equation. It's convenient to
restrict  the $u_{\bf{k}}(t,{\bf{x}})$ to the interior of a three
dimensional torus of side $L$ (i.e., choose periodic boundary
conditions). Then
\begin{equation}
u_{\bf{k}}(t,{\bf{x}}) = (2L^3 \omega)^{-1/2}e^{i{\bf{k}}.{\bf{x}}
- i\omega t}
\end{equation}
where
\begin{equation}
k_i = 2\pi j_i/L    \,\,\,\, j_i=0,\pm 1,\pm 2... \,\,\, i=1,2,3.
\end{equation}

It is possible to show that the interaction Hamiltonian defined by
Eq.(\ref{33}) is equivalent to the interaction hamiltonian given
by $H_{I}=m(\tau)\varphi(x(\tau))$. This model is known as the
Unruh-Dewitt detector. The detector is an idealized point-like
object with internal degrees of freedom defining two energy
levels. Different coupling between the field and the two-level
system was analyzed by Hinton \cite{hinton}.

\end{appendix}

\end{document}